\def\year{2022}\relax
\documentclass[letterpaper]{article} 
\usepackage{aaai22}  
\usepackage{times}  
\usepackage{helvet}  
\usepackage{courier}  
\usepackage[hyphens]{url}  
\usepackage{graphicx} 
\usepackage{natbib}  
\usepackage{caption} 
\DeclareCaptionStyle{ruled}{labelfont=normalfont,labelsep=colon,strut=off} 
\usepackage{algorithm}
\usepackage{algorithmic}
\usepackage{tabularx}
\usepackage{xcolor}

\usepackage{newfloat}
\usepackage{listings}
\floatstyle{ruled}
\newfloat{listing}{tb}{lst}{}
\floatname{listing}{Listing}

\usepackage{subfig}

\usepackage{xcolor}

\title{Link Me Baby One More Time: Social Music Discovery on Spotify}
\author {
    Shazia'Ayn Babul\textsuperscript{\rm 1,2,3}\footnote{This work was completed at an internship at Spotify.}\footnote{These authors contributed equally.},
    Desislava Hristova, \textsuperscript{\rm 3}$^\dagger$,
    Antonio Lima \textsuperscript{\rm 3},\\
    Renaud Lambiotte \textsuperscript{\rm 1, 2},
    Mariano Beguerisse-D\'iaz \textsuperscript{\rm 3,1}
}
\affiliations {
    \textsuperscript{\rm 1} Mathematical Institute, University of Oxford, Oxford, UK\\
    \textsuperscript{\rm 2} The Alan Turing Institute, London, UK\\
    \textsuperscript{\rm 3} Spotify, London, UK\\
    shazia.babul@maths.ox.ac.uk, desih@spotify.com, antoniol@spotify.com,
    renaud.lambiotte@maths.ox.ac.uk,
    marianob@spotify.com
}

\begin{document}

\maketitle

\begin{abstract}
We explore the social and contextual factors that influence the outcome of person-to-person music recommendations and discovery. Specifically, we use data from Spotify to investigate how a link sent from one user to another results in the receiver engaging with the music of the shared artist. We consider several factors that may influence this process, such as the strength of the sender-receiver relationship, the user's role in the Spotify social network, their music social cohesion, and how similar the new artist is to the receiver's taste.
We find that the receiver of a link is more likely to engage with a new artist when (1) they have similar music taste to the sender and the shared track is a good fit for their taste, (2) they have a stronger and more intimate tie with the sender, and (3) the shared artist is popular amongst the receiver's connections.
Finally, we use these findings to build a Random Forest classifier to predict whether a  shared music track will result in the receiver's engagement with the shared artist. This model elucidates which type of social and contextual features are most predictive, although peak performance is achieved when a diverse set of features are included. 
These findings provide new insights into the multifaceted mechanisms underpinning the interplay between music discovery and social processes. 

\end{abstract}
\section{Introduction}

Music is a staple of human society, functioning as a means for social bonding and cohesion~\cite{savage_loui_tarr_schachner_glowacki_mithen_fitch_2021}. Online streaming platforms have changed the ways that people discover, listen, and share music with each other~\cite{wang2023exploring, mok2022dynamics}.  We are interested in understanding the social aspect of music sharing, or how person-to-person recommendations facilitated by established social relationships lead people to discover and appreciate new music.  

Spotify is an audio streaming platform where users can listen to music and share it with others. Social music sharing is facilitated by a function that allows users to send Spotify content links over various social media and messaging platforms.
When these person-to-person recommendations are successful, the receiver engages with 
the shared artist's music, potentially becoming a new fan over time.
We elucidate the social and contextual factors that lead to a successful recommendation (or ``activation"), creating a predictive model for short-term engagement, and examining the spread of new music discoveries over the Spotify social network.

Models of diffusion 
processes over social networks often are inspired by epidemic spread models, by relating the activation probability to the social network structure and the speed of the spread~\cite{newman2002spread}. These models can often be applied to understand information diffusion in an online context~\cite{gruhl2004information}. Other social processes require 
reinforcement to spread effectively, as in Granovetter's threshold model, where activation probability increases as a function of number of previously activated connections~\cite{granovetter1978threshold}. The successful adoption of behaviours requiring social reinforcement, called complex contagion, is deeply impacted by network topology, and spread more easily over clustered networks~\cite{centola2010spread}. Empirical evidence for complex contagion has been found to describe behaviours from information sharing on social media~\cite{monsted2017evidence}, political campaign donations~\cite{traag2016complex}, person-to-person product recommendations~\cite{leskovec2007dynamics}, and even the decision to exercise~\cite{aral2017exercise}. 

In these works, there is often no evidence of “contact” events between individuals, which must instead be inferred from the network topology and sequence of activations.  In the absence of contact event information, it can be difficult to understand which spreading process is occurring. Several works have proposed methods for inferring the process type and parameters using network topology and the observable activations~\cite{cencetti2023distinguishing, dutta2018bayesian}. Others have taken a global approach, to understand how diffusion processes can lead to cascades on a social network~\cite{sun2009gesundheit, lerman2010information}.  

There have been few works applying spreading models specifically to the musical context. A recent study showed how live music events generated increased listenership amongst concert attendees and their friends \cite{TERNOVSKI2020144}. Another observational study compares song download time-series to epidemic spread curves to identify mechanism that contribute to the spread of a song~\cite{rosati2021modelling}. This macro-level approach is common in other works applied to different contexts on empirical datasets, seeking to observe evidence of complex contagion on social networks using just the adoption curves \cite{fink2016investigating}. 

In contrast to other related works, in this paper we have evidence of the sharing events between users, in the form of link shares, as well as information about both senders and receivers, including prior listening history, relationship history, and role in the social network. This information enables us to investigate how these social and contextual factors effect the probability of a successful social recommendation. 

Our work here, like that of \cite{leskovec2007dynamics}, investigates person-to-person recommendations. These processes are distinct from those studied in the majority of works on social cascades which focus on information propagation, such as cascades over social media. In these cases, the action of “adopting” the post is intertwined with the action of socially spreading it - by clicking to re-share. In our case, we are instead interested in the change of behaviour of an individual after receiving a recommendation - the two actions, adoption and propagation are therefore distinct, and sharing requires a higher level of individual effort. In \cite{cheng2018diffusion}, the authors characterise different diffusion protocols based on level of individual effort and social cost of not participating.  Sharing music to a friend has a high individual effort, while there is a low social cost to engaging more with the shared artist. The authors assert that such protocols spread more effectively with social reinforcement, and over stronger social ties. Peer influence is generally more effective over stronger and reciprocal ties \cite{almaatouq2016role}, and empirically found to be the case in an experiment studying political mobilisation on Facebook~\cite{bond201261}. 

Our main hypotheses about the social aspects of music discovery via link shares on Spotify are that:

\begin{itemize}
    \item (H1) Post-share engagement of a receiver with the shared artist is related to music taste similarity.
    \item (H2) Post-share engagement of a receiver is more likely to happen along stronger, established ties and through more intimate modes of contact such as direct messaging.
    \item (H3) Post-share engagement of a receiver depends on music social cohesion among the receiver's social contacts.
\end{itemize}

Music taste similarity (H1) relates to whether the sender and receiver share similar tastes and the shared track is similar to the receiver's historical listening taste. Our analysis of (H2), which is inspired by the above-mentioned literature on social influence, explores various definitions of social tie strength, such as the number of social interactions and whether the link share was a direct message or shared more widely on social media. Inspired by complex contagion literature, we look at the proportion of contacts who are already engaged with the shared artist in the receiver's social network (social cohesion), as well as the degree of network clustering, and the social group overlaps between the sender and receiver (H3).


Our paper is organised as follows. We first describe our dataset, consisting of track share events, the Spotify social network, and artist and user attributes. We describe the method we use to sub-sample the data, and to classify a successful person-to-person recommendation from post-share engagement. In the results section, we investigate the three hypotheses by analysing the relationship between future engagement probability and relevant features. Finally, we build a Random Forest model to predict short-term engagement following a link share event, and explore the different groups of features that are most vital for the performance of the model.  

\section{Data and Methods}
Spotify has over 574 million users, and over 100 million available tracks~\cite{Spotify_2023}. The platform also offers a social aspect; users have various ways of interacting with others on Spotify, from following the activity of their friends, sharing music, to collaboratively building playlists together. For this study, we use datasets describing the listening and sharing history of users on the platform, as well as their social activity with other users.

\begin{figure}
\centering
\includegraphics[width=\linewidth]{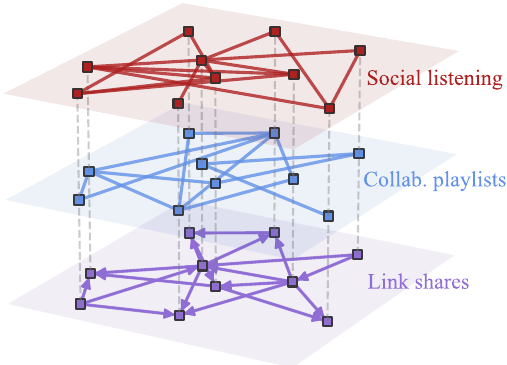}
\caption{A synthetic network representing the structure of Spotify social network. The Spotify social network is a multiplex network, each layer corresponding to a different social interaction. Social listening sessions and collaborative playlists represent undirected layers, while link sharing is a directed layer.}
\label{fig:multilayer}
\end{figure}



\subsection{Spotify Social Network}
The Spotify social network comprises its users (nodes) and their interactions on the platform (edges). Each edge has an associated timestamp, indicating when an interaction occurred. This network is multiplex, each layer representing a different type of interaction; collaborative playlists, social listening sessions and link shares. 
A collaborative playlist is a special type of playlist that can be modified by a group of users. Although these are group interactions, we represent them as multiple pairwise interactions (edges) between each pair of collaborating users. The edges representing these interactions are undirected, as they are a reciprocal interaction. A social listening session allows multiple users to add songs to a real-time queue, listening to the same music either remotely or together. Like collaborative playlists, these are reciprocal interactions, and the edges are undirected. A link share is a link to a track, album or artist sent from one user to another via a messaging or social media platform. This interaction is directed, and is represented in the network by an edge from the sender to the receiver. 

Figure \ref{fig:multilayer} gives a depiction of this layered structure on  a synthetic network. Each layer can be described by an adjacency matrix $L$, an $N \times N$ matrix where $N$ is the number of users. The values of the matrix $L_{ij}$ are the number of interactions occurring from user $i$ to user $j$. In the case of an undirected layer, where the interactions are reciprocal, the matrix is symmetric and $L_{ij} = L_{ji}$.

The resulting social network has three layers: 

\begin{itemize}
    \item (1) Social listening sessions, an undirected layer with adjacency matrix $L_{s}$. 
    \item (2) Collaborative playlists, an undirected layer with adjacency matrix $L_{c}$. 
    \item (3) Link sharing, a directed layer with adjacency matrix $L_{l}$.
\end{itemize}

When aggregating the adjacency matrices of all the layers as $L_{s} + L{c} + L_{l}$, the network has $\sim$2.2B edges between $\sim$518M users (nodes). We specify different aggregations of these layers in the results sections, depending on the metric we are computing.

\subsection{Link Sharing Events}
To track the spreading of behaviour on the network, we focus on tracks shared between pairs of users, which can happen over various social media or messaging platforms. These shares give a direct signal of music spreading between people, which we can use to understand how the music discovery process on Spotify is influenced by social factors.

We only consider ``discovery” share events where the receiver has not listened to the artist before and we filter the data to include only link shares of tracks for which 1) the link was opened (comprising 50\% of all shares), and 2) the receiver's Spotify account registered a playback event of the shared content for at least 30 seconds (64\% of all opened shares). This ensures that the receiver did indeed come into contact with the music that was shared. To account for popularity bias, we randomly sample up to 1M share events stratified by artist across various popularity bins. Our final sample consists of $\sim$2.72M events across 3,102 distinct artists between April-June 2023. Each link sharing event contains: 
\begin{itemize}
    \item Anonymised sender and receiver identifiers.
    \item Track and album identifiers, and the primary artist identifier.
    \item Artist popularity rank at time of share.
    \item Album release age from time of share.
    \item Social media application type (e.g. WhatsApp).
    \item Share timestamp and share open timestamp.
\end{itemize}
The identifiers and timestamps define the link sharing events, which are the basis of our analysis. The artist popularity, album age and application type are features in our analysis and predictive model, also described in Table~\ref{features_described}.





\begin{figure}
\centering
\includegraphics[width=\linewidth]{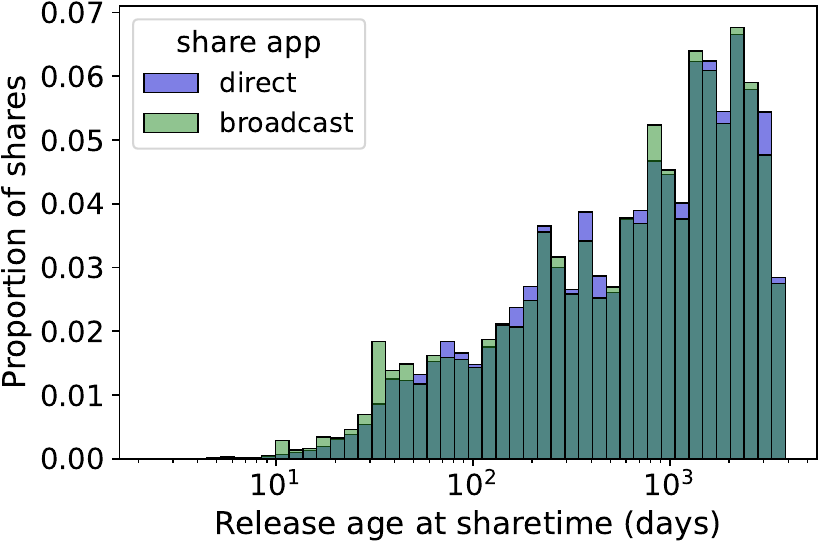}
\caption{Distribution of the album release age at share-time for our link sharing events, separated by broadcast and direct application types.}
\label{fig:release}
\end{figure}

We split the social media application type into two categories, direct or broadcast, depending on whether they represent one-to-one or one-to-many modes of communication.
The applications we identify as direct are: WhatsApp, Facebook Messenger, SMS, Line, Instagram Direct, and Samsung Messenger. The applications we identify as broadcast are: Instagram Stories, Facebook Feed, Facebook Stories, and X, formerly Twitter.  Some sharing events have an unknown destination platform or fit into neither category. These events are excluded from the analysis where share category is relevant. Note that the applications we identify as one-to-one can also be used to message group chats, but the key distinction here is in the magnitude of number of people receiving the link. Figure \ref{fig:release} gives the distribution of album release age at share-time for our link sharing event sample, separated by direct and broadcast integration. We can see that older tracks are more likely to be shared because there is more old content than new on the platform.


\subsection{Music Vector Embeddings}
Some works on spreading process treat each individual in the social network as identical, with identical susceptibility for activation~\cite{bailey1975mathematical}. Here, we use each user's prior listening history to investigate the relationship between music discovery, music taste similarity, and homophily, or the inclination for people to connect with others who are similar to them.  

To this end, we use vector representations of Spotify users and content in latent space, which are pre-trained and available internally at Spotify. These vector embeddings are computed daily on Spotify by training the word2vec algorithm \cite{mikolov2013efficient} on user generated playlists, with the playlists as the ``documents" and the tracks as the ``words". This process results in 80-dimensional embedding vectors for each track, where tracks that are often co-listed in playlists appear close together in the embedding space. Users are also embedded in this space by taking the average of vectors over all tracks they have listened to over a certain time window. For more information on how these vector embeddings are generated, see~\cite{anderson2020algorithmic}. The information captured by the embedding space methodology has been reported to be a good representation of the way that users engage with music~\cite{mok2022dynamics, anderson2020algorithmic}. We use vectors computed at the beginning of our analysis period for each user and track entity in our sample, to ensure consistency. The vectors can be used to understand the taste similarity of two users, or between a user and track. We define this as the cosine similarity between the two vectors. 

\subsection{User-Artist Engagement}

\begin{figure}[t!]
\centering
\subfloat[\centering]{{\includegraphics[width=\linewidth]{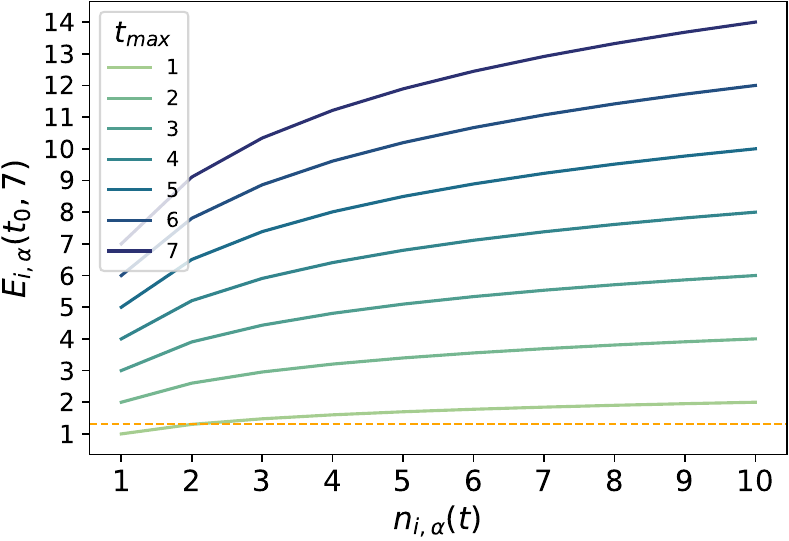}}}%
\qquad
\subfloat[\centering]{{\includegraphics[width=\linewidth]{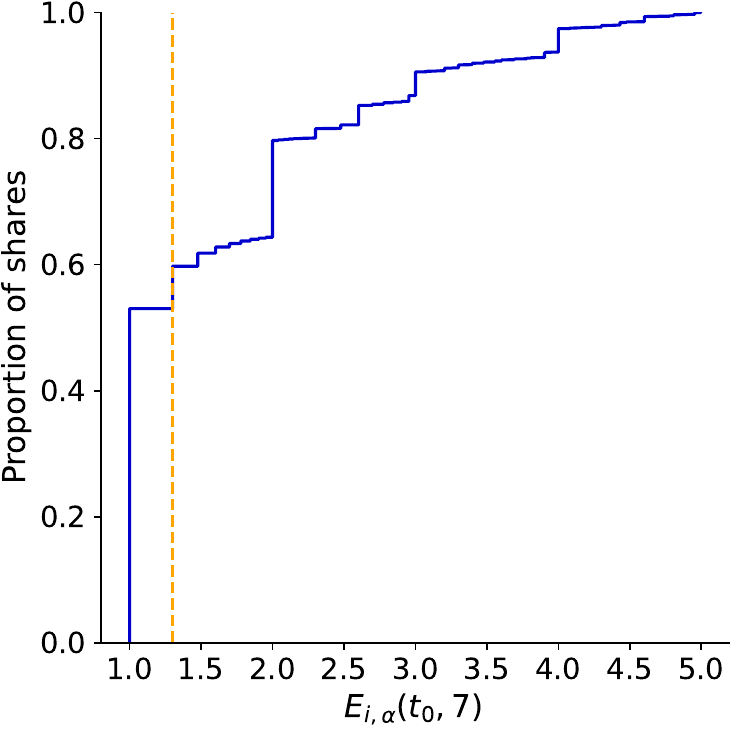}}}%
\caption{User-artist engagement threshold choice. (a) User-artist engagement curves as a function of the number of tracks $n_{i,\alpha}(t)$ for different number of days listened $t_{\max}$ within a 7 day period. 
(b) Cumulative distribution function (ecdf) of the link shares and the associated receiver-artist engagement 7 days post share ($E_{i, \alpha}(t_{0}, 7)$. The orange line indicates our chosen threshold for a successful share event.}
\label{fig:engagement}
\end{figure}

To define the success of a share, we are interested in a user's engagement with an artist before and after a share event. We denote the number of unique tracks by the artist $\alpha$ that the user $i$ streams for longer than 30s on a particular day $t$ as $n_{i,\alpha}(t)$.  We then define the user-artist engagement as $e_{i,\alpha}(t) = \log_{10}(n_{i,\alpha}(t)) + 1$, if $n_{i,\alpha}(t) \geq 1$, and 0 otherwise.  User-artist engagement captures the intensity of the user's engagement with the artist each day, which we aggregate over time as:
\begin{equation}
   \label{eqn:engagement}
    E_{i,\alpha}(t_0, K) = \sum_{t=t_0}^{t_0+K} e_{i,\alpha}(t),
\end{equation}
where $t_0$ is the day when the link share is opened, and $K$ is the number of days over which we aggregate, which can be either positive or negative indicating aggregation over time before or after $t_0$ respectively. There are several choices that can be made to measure user-artist engagement. For example, the total number of streams over a window of time gives a broad indication of interest, the number of unique tracks listened to over a window of time gives an indication of deep interest, or the number of days listened to an artist, which is a longitudinal metric indicative of loyalty.  Our choice of $E$ in Eq.~(\ref{eqn:engagement}) balances all three by looking at the unique number of daily tracks listened over a time window. The deeper (more daily tracks) and extended (more days) the listening is, the higher the engagement. The advantage of using this metric, instead of simply counting the number of unique tracks that the user listens to over the same period of time, is that $E_{i,\alpha}(t_0, K)$ is higher if the number of listening events are dispatched over a larger number of days. For example, Fig.~\ref{fig:engagement}~(a) shows the profile of $E$ for a varying number of daily unique tracks listened $n_{i,\alpha}(t)$ and the total number of days listened $t_{\max}$ within a 7 day period. As we increase the number of days of engagement and the number of tracks listened, we define different curves for the engagement which reward high and repeat engagement. This behaviour is desirable when trying to capture the development of a habitual relationship such as becoming a fan of an artist. Thus, we capture the delineation between a state of longer-term engagement, and an intense, but time-limited, fixation with the artist. We use this metric to define the past engagement of a sender and the future engagement of a receiver with the shared artist.

We define the success of a share event as the binary outcome of the receiver's future engagement with the shared artist. Figure~\ref{fig:engagement}~(b) gives the cumulative distribution function of $E_{i,\alpha}(t_0, 7)$, the receiver-artist 7 day post-share engagement. We observe that $52.8\%$ of shares result in an engagement of 1.0, indicative of the user only listening to the shared track once in the 7 days post-share. Consequently, we define an engaged receiver as anyone who has listened to two or more tracks by the artist in the 7 days following the share,
this is equivalent to a threshold of $E_{i,\alpha}(t_0, 7) > 1.3$, since $\log_{10}{2+1} = 1.3$. This threshold is shown in Fig.~\ref{fig:engagement} (b) by the dotted orange line.
Intuitively, exploring an artist's other tracks and returning to the artist's music after the initial share is a good sign that the user is engaging with the artist and the recommendation is successful, and so this choice of threshold identifies receiving users who are beginning to form an attachment with the artist in the short-term and are becoming potential fans. By this definition, $47.2\%$ of link shares result in successful engagement, and fairly balanced split between the link share events. Note that there are other choices that could be made to define a successful share. In particular, here we are interested in behaviour change on a short-term timescale directly following a share.

\section{Results}
We evaluate the three hypotheses set out at the beginning of the paper by examining the probability of engagement $p(\mathrm{engaged})$, i.e. the proportion of link share events where the receiver's engagement is $E_{i,\alpha}(t_0, 7) > 1.3$ out of all share events, with respect to: the music taste similarity of the pair of sharing users and the shared track (H1), their relationship strength (H2), the social cohesion of their neighbourhood with respect to an artist (H3). We then construct and evaluate a predictive model of new listener engagement, using a broader set of social and contextual features.

\begin{figure}
\centering
\includegraphics[width=\linewidth]{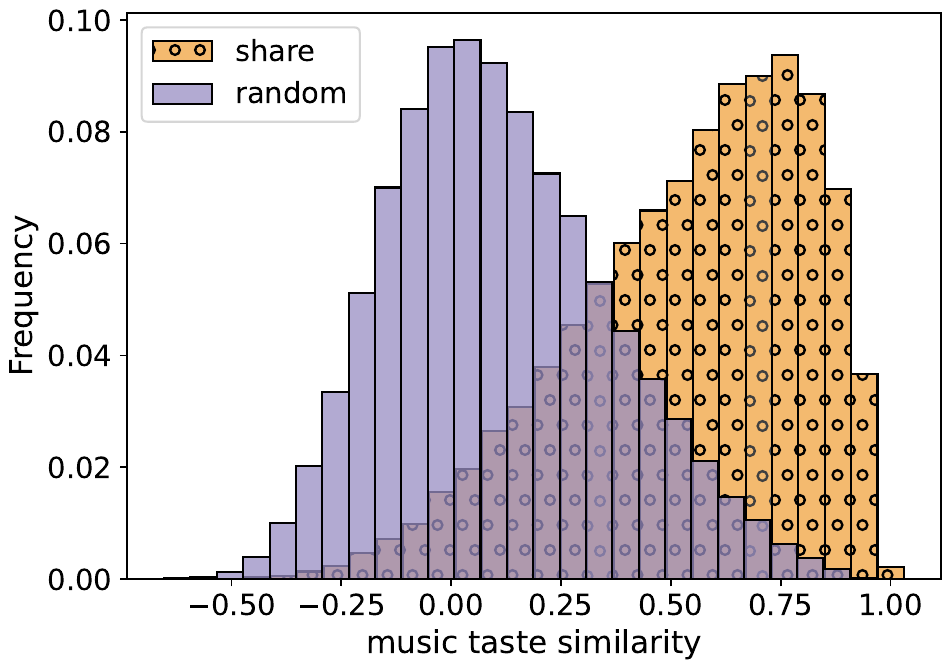}
\caption{Distributions of music taste similarity for pairs of users who shared music in our sample (share) and a random permutation of those pairs (random).}
\label{fig:homophily}
\end{figure}

\subsection{Music Taste Similarity}
Social networks often exhibit homophily, the phenomenon that social ties tend to form between similar individuals~\cite{mcpherson2001birds}. For this reason, it is often difficult to determine whether correlated patterns of behaviours between connected individuals have a causal influence (social contagion) or arise as a result of homophily. Several works have proposed methods to separate them, though the problem remains difficult without longitudinal data~\cite{vanderweele2013social, aral2009distinguishing}.

\begin{figure*}
    \centering
    \subfloat[\centering]{{\includegraphics[width=0.45\linewidth]{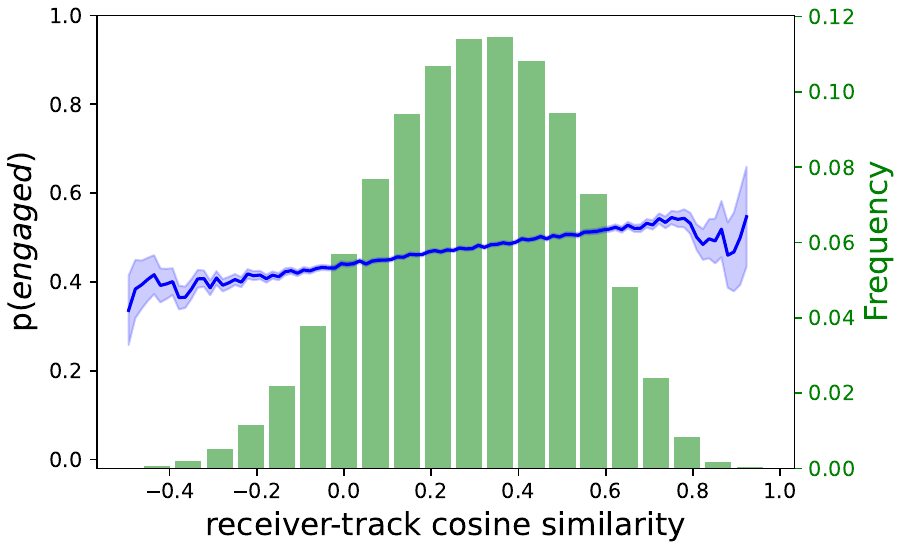}}}%
    \qquad
    \subfloat[\centering]{{\includegraphics[width=0.45\linewidth]{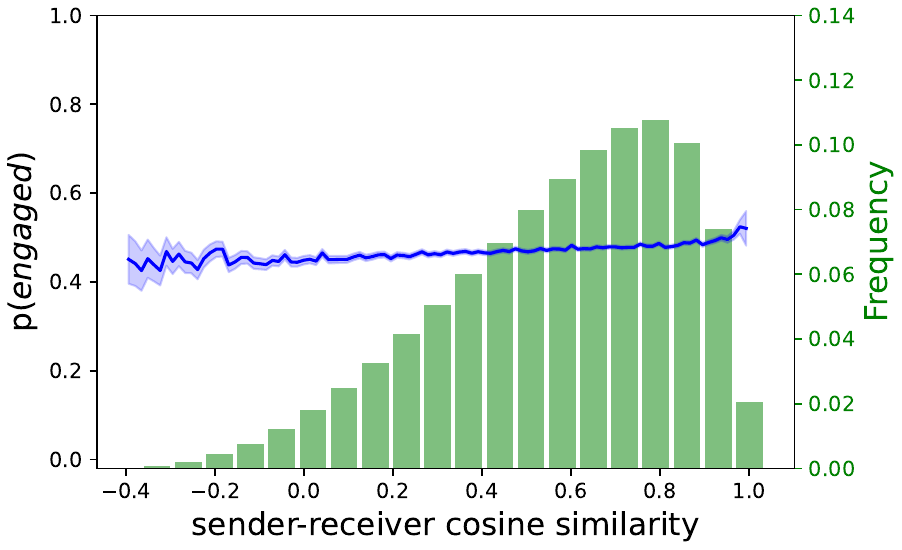}}}%
    \caption{Analysis of music taste similarity. The blue line shows the probability of becoming engaged with an artist as a function of the feature, where the shaded area is the $95\%$ confidence interval (main y-axis). The histograms give the distribution of the feature in our sample in the second y-axis (Frequency).  
    } %
    \label{fig:taste_affinity_figs}%
\end{figure*}

We measure music taste similarity as the cosine similarity of the music vectors described in Data and Methods. We first observe the level of homophily present in the sample of people sharing links, whether successful or not. We randomly shuffle the sender in user pairs in our sample, creating random permutation pairs of the social network, and take the user pair cosine similarity of their taste vectors. Figure \ref{fig:homophily} compares the distribution of the user pair cosine similarities between random pairs, and between the pairs who have shared and opened links, where we note that the distribution of user pair cosine similarities for users involved in link sharing is skewed towards higher similarity with a mean of 0.54, compared to the random pairs with a mean of 0.11. The difference in distributions between the two groups is statistically significant as confirmed by the Kolmogorov-Smirnov test ($D=0.44, p < 0.01$). This gives some indication of the level of homophily in the Spotify social network; either users are more likely to become friends with other users who share similar tastes, or they are more likely to share with their friends who have similar tastes.

For this reason, it is difficult to determine whether the engagement we observe following a share is due to homophily or social contagion. However, while we may not be able to distinguish between the two mechanisms at play, we can still investigate the social and contextual factors that lead to engagement as a new listener of the shared artist. We first test the hypothesis (H1) that the probability of engagement will increase when users have similar taste in music (sender-receiver cosine similarity), or when the shared track is similar to the listening history of the receiver (receiver-track cosine similarity). The results are given in Fig.~\ref{fig:taste_affinity_figs} (a), showing the distribution of receiver-track cosine similarity in our sample, and $p(\mathrm{engaged})$ as a function of the similarity. Similarly, Figure \ref{fig:taste_affinity_figs} (b) shows the results for sender-receiver cosine similarities. We observe that while (H1) is confirmed, there is a stronger effect for the receiver-track cosine similarity; $p(\mathrm{engaged})$ increases linearly with cosine similarity. 

\subsubsection{Sender-Artist Engagement}
\begin{figure}
\centering
\includegraphics[width=0.9\linewidth]{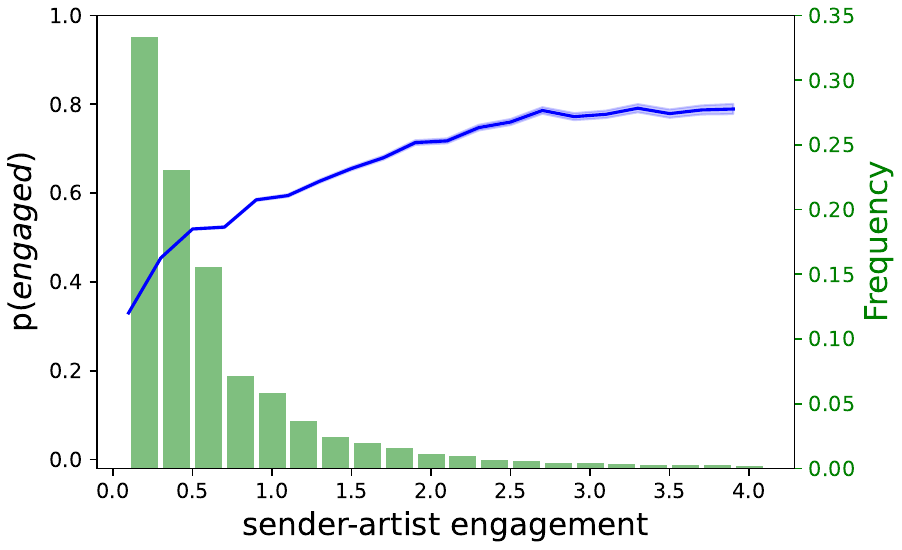}
\caption{Sender-artist engagement aggregated 7 days prior to share. The blue line gives the probability of receiver engagement (main y-axis) as a function of the feature, where the shaded area is the $95\%$ confidence interval.  The histogram (second y-axis in green) gives the distribution of the feature in our sample. }
\label{fig:sendartaf}
\end{figure}

We also investigate the effect of the sender-artist engagement for the artist on $p(\mathrm{engaged})$. We measure this by taking the 7 day aggregate user-artist engagement (described in Data and Methods) for the week prior to the share, which is a signal for how engaged the sender was with listening to the artist, prior to sharing their music. This corresponds to Equation \ref{eqn:engagement}, where $i$ is the sender, $t_{0}$ is the day of the share, and $K = -7$. This feature can be thought of as a proxy for the strength of the sender's recommendation. Users have the ability to add messages when sharing a Spotify link, and while we do not have access to this textual data, it is possible that the associated message might be more emphatic when the sender is more engaged with the artist. Note that the sender-artist engagement can be 0 if the sender sends a track to someone without listening to it first, as these are separate functions on the Spotify interface. These results are shown in Fig.~\ref{fig:sendartaf}. The probability of engagement of the receiver increases with the sender-artist engagement.

\subsection{Tie Strength Effects}
\begin{figure*}[t]
    \centering
    \subfloat[\centering]{{\includegraphics[scale=0.5]{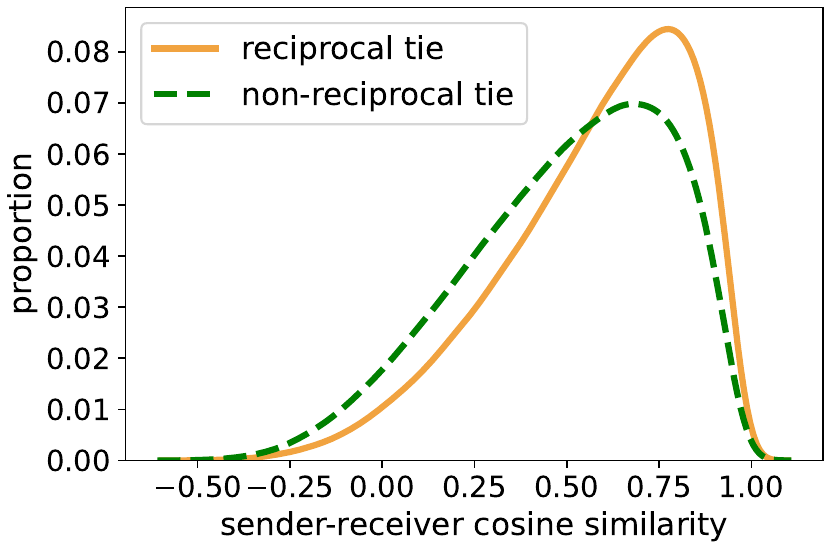}}}%
    \subfloat[\centering]{{\includegraphics[scale=0.5]{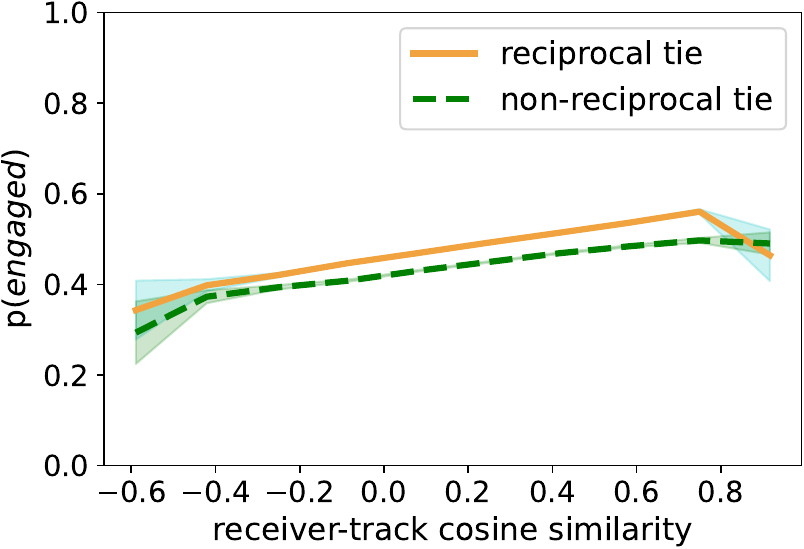}}}%
    \qquad
    \subfloat[\centering]{{\includegraphics[scale=0.5]{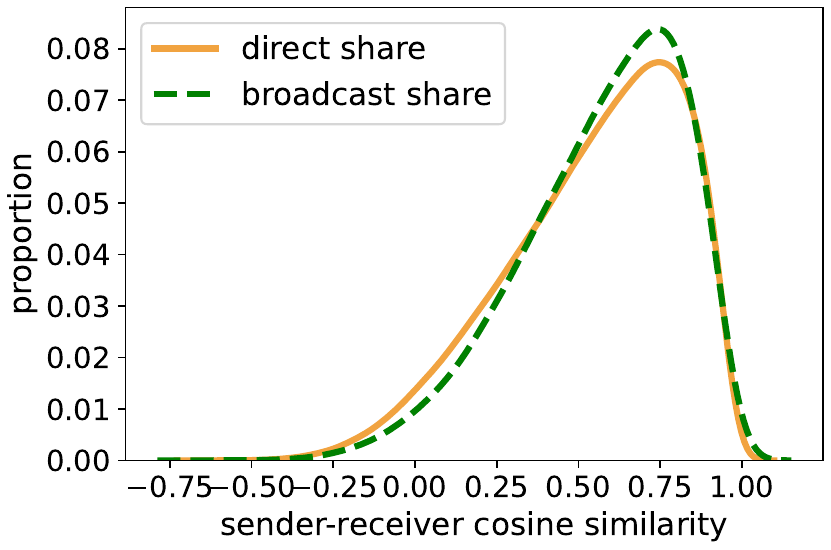}}}%
    \subfloat[\centering]{{\includegraphics[scale=0.5]{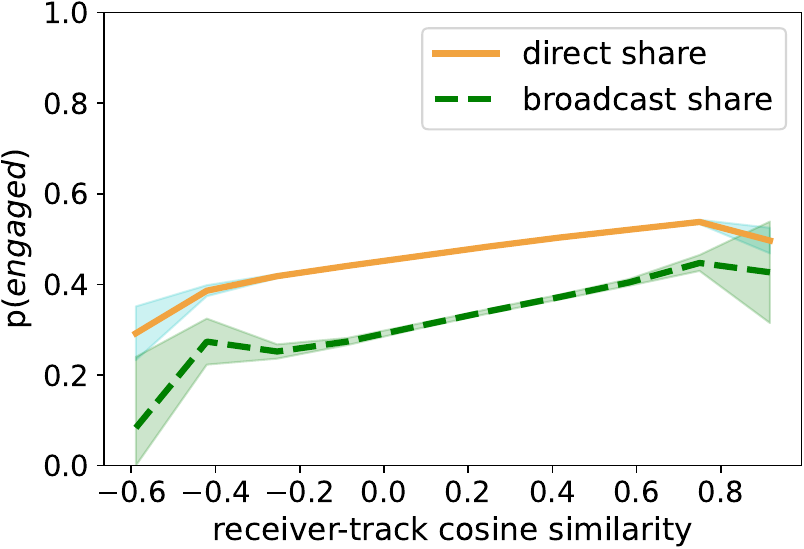}}}%
    \caption{Effects of reciprocity (a-b) and share media type (c-d) on the probability of becoming engaged with the artist: (a) Distribution of sender-receiver cosine similarities in our sample, separated by reciprocal and non-reciprocal tie. (b) Probability of engagement following link share as a function of receiver-track cosine similarity, separated by reciprocal and non-reciprocal tie.
    (c) Distribution of sender-receiver cosine similarities in our sample, separated by direct or broadcast application. (d) Probability of engagement following link share as a function of receiver-track cosine similarity, separated by direct or broadcast application.}%
    \label{fig:tie}%
\end{figure*}

As suggested in \cite{almaatouq2016role} and \cite{bond201261}, stronger social ties are more effective at spreading behaviours. We explore the hypothesis (H2) that the probability of engagement $p(\mathrm{engaged})$ will increase with stronger and more intimate ties, as represented by the tie strength from the social network. We measure tie strength in three ways; reciprocity, the mode of social media application used to share the track, and sum of social interactions. Reciprocity, is a boolean variable that indicates whether the receiver ($i$) has ever shared a link with the sender ($j$), prior to share time. Specifically, we measure if $L_{l;ij} > 0$. Figure \ref{fig:tie} (a) shows the distribution of sender-receiver cosine similarities over all share events, separated by reciprocal and non-reciprocal ties. For reciprocal links, the mean sender-receiver cosine similarity is 0.58, compared to 0.5 for non-reciprocal links. The difference in distributions between the two groups is statistically significant as confirmed by the Kolmogorov-Smirnov test ($D=0.12, p < 0.01$). This indicates that users connected by reciprocal ties tend to have a greater taste similarity than those who do not. Figure \ref{fig:tie} (b) shows the probability of becoming an engaged receiver, $p(\mathrm{engaged})$, along a reciprocal and non-reciprocal link, holding constant the receiver-track cosine. Even accounting for the taste of the receiver, the probability of future engagement is more likely along a reciprocal tie. There is noise at very large and very small receiver-track cosine similarities, where there is little data available.

The second measure we investigate here is the mode of social media application, which indicates whether the application used to share the link represents a one-to-one (direct) or one-to-many (broadcast) mode of communication. Although this is not strictly a measure of tie strength, it is intuitive that one-to-one communication methods (like Whatsapp or Facebook Messenger) are more intimate, as compared to one-to-many (posting an Instagram story).  In our analysis, we exclude events for which the mode of social media application is unknown or ambiguous, keeping only known direct (66\%) or broadcast (5\%) share events. 

Figure \ref{fig:tie} (c) shows the distribution of sender-receiver cosine similarities over all share events, separated by ties that represent direct or broadcast platform share events. For the users sharing via direct social media applications, the mean user-user cosine similarity is 0.54, compared to the 0.57 for users sharing via broadcast methods. The difference in distributions between the two groups is statistically significant as confirmed by the Kolmogorov-Smirnov test ($D=0.03, p < 0.01$). This indicates that users selectively engage with music shared by friends with similar taste on social media.
Figure \ref{fig:tie} (d) shows the probability of becoming engaged with the artist for share events via direct and broadcast mediums, again holding constant the user-track cosine similarity. Accounting for the taste of the receiver, the future engagement with an artist is more likely to occur when the mode of communication is a direct one, even when the track is not as good a match for the receiver's taste. Note that the probability of engagement goes almost to 0 for broadcast application types at large receiver-track cosine similarities.

\begin{figure}
    \centering
    \includegraphics[scale=0.5]{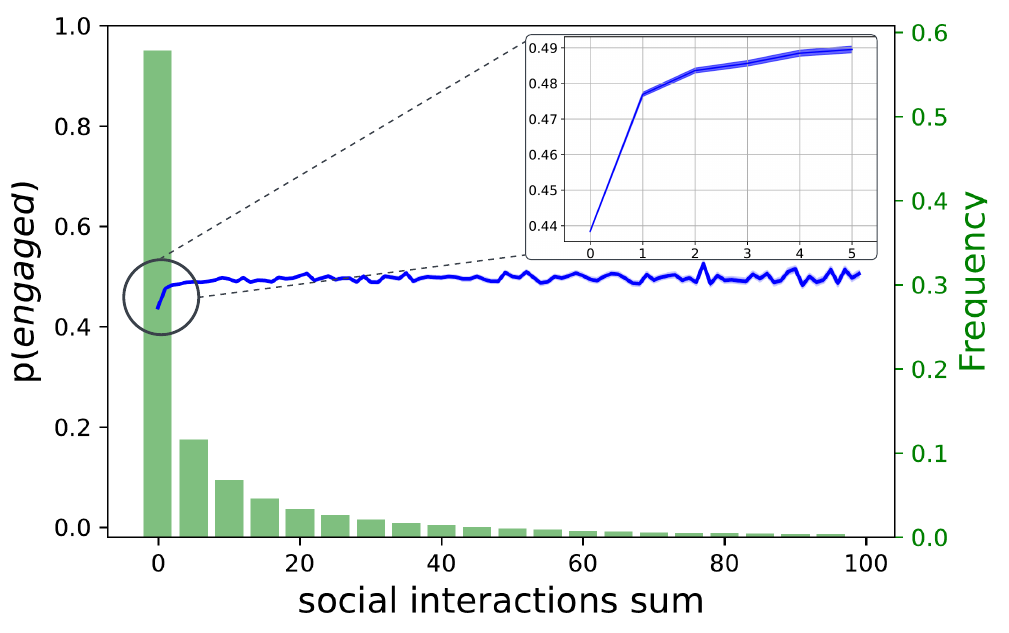}
    \caption{Analysis of sum of social interactions. The histogram gives the distribution of the feature in our sample. The blue line gives the probability of engagement as a function of feature, where the shaded area is the $95\%$ confidence interval. The inset gives a closer look at the probability of engagement at the lower end of total interactions. 
    }%
    \label{fig:priorshare}%
\end{figure}

The final measure of tie strength, sum of social interactions, the number of times that the sender ($j$) and receiver ($i$) have interacted before the share on Spotify. This includes link shares (from the sender to receiver or from the receiver to the sender), social listening sessions and blends. The sum of social interactions is equivalent to $L_{s;ij} + L_{c;ij} + L_{l;ij} + L_{l;ji}$, where we have included the link sharing in both directions. This measure gives a sense of whether the social tie is new, or previously established. Figure \ref{fig:priorshare} shows the distribution of the sum of social interactions over the pairs of users in our sample, as well as the probability of engagement along a link as a function of the sum of social interactions. The probability of engagement jumps from 0.44 to 0.48 just after one previous interaction and then gradually up to 0.5, until 5 interactions. After 5 interactions, the probability of engagement remains stable. This indicates that while engagement is more likely to happen along an established tie, there is a ``saturation point" at which more previous interaction does not increase the probability. 


\begin{figure*}[th!]
    \centering

    \subfloat[\centering]{{\includegraphics[scale=0.4]{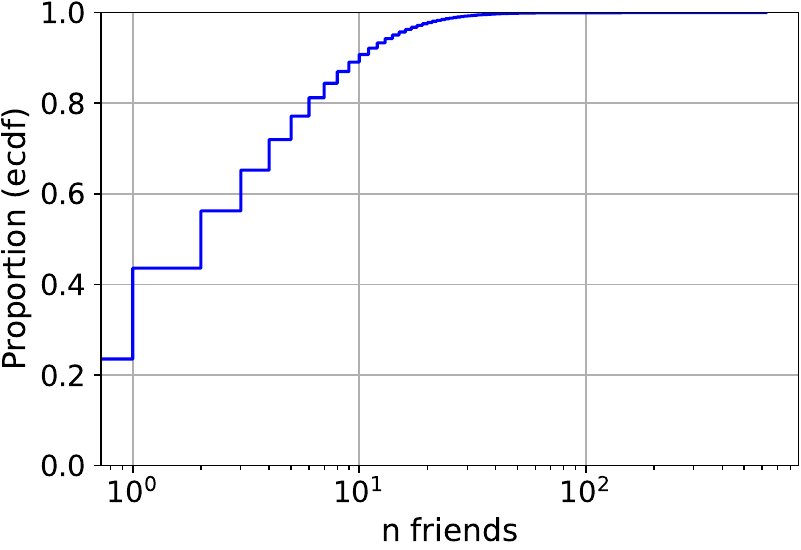}}}%
    \qquad
    \subfloat[\centering]{{\includegraphics[scale=0.4]{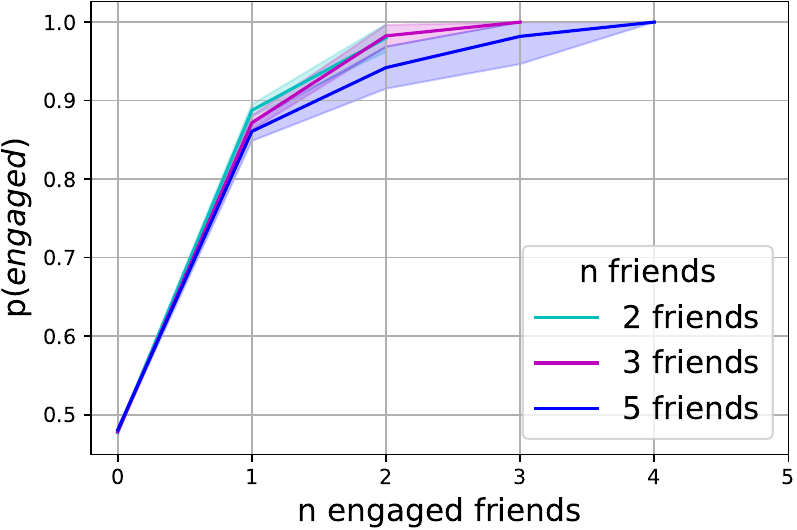}}}%
    \subfloat[\centering]{{\includegraphics[scale=0.4]{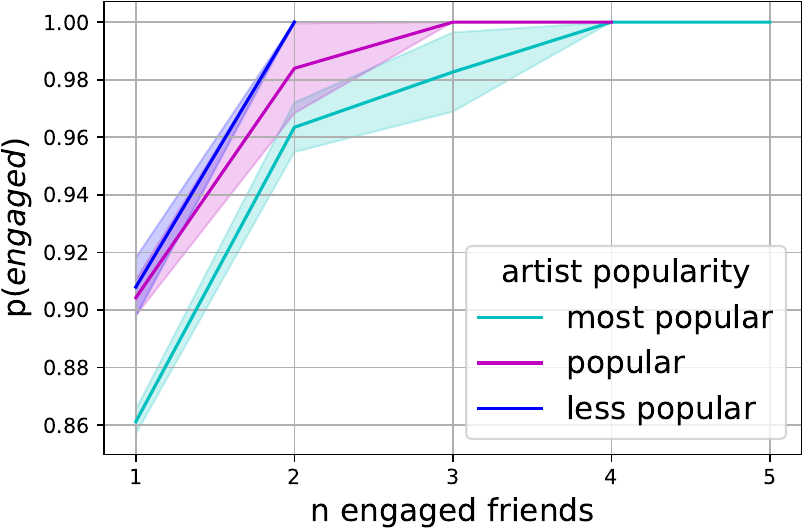}}}%
    \caption{Analysis of social cohesion features. (a) Distribution of number of friends for all receiving users in our sample. (b) Probability of becoming engaged versus number of friends who are already engaged with the artists, for receivers with different numbers of friends. (c) Probability of engagement versus number of friends who are already engaged, stratified by artist popularity.}%
    \label{fig:pfan}%
\end{figure*}

Note that we do not assume that these three measures of tie strength are independent of each other. For example, reciprocal ties and direct shares tend to have more prior social interactions in the social network. Direct ties have on average a sum of social interactions of 24 compared to 6 for broadcast, while reciprocal ties have an average of 42 social interactions, compared to 3 for non-reciprocal ties. We include all three features in this analysis as they represent different aspects and granularity of tie strength, and indeed show different relationships with the probability of post-share engagement.

\subsection{Social Cohesion}

Next, we explore the effects of social cohesion. These results investigate how the level of engagement with the shared artist in the receiver's neighbourhood of contacts affects the probability of becoming engaged as a receiver of a share link (H3). We define the receiver's ``friends" as anyone with whom they have a reciprocal relationship in the social network. This includes social listening sessions, collaborative playlists, or link sharing (must be in both directions). More specifically, the receiver $(i)$ is friends with a user $(u)$ if $L_{s;iu} > 0$, or $L_{c;iu} > 0$, or $L_{l;iu} > 0$ and $L_{l;ui} > 0$. Figure \ref{fig:pfan} (a) shows the distribution of number of friends over receivers in the sample. The average number of friends is 3.95, but the distribution is skewed to the left, most users have 0 or 1 friend. A receiver's friend is considered to be engaged with the shared artist if they have a user-artist engagement indicative of listening to at least one track per day on average by the shared artist in the six months prior to the analysis period. This is equivalent to the condition $E_{i, \alpha}(t_0, 180) > 180$, where $i$ is the friend user, $\alpha$ is the shared artist and $t_0$ is the start of the analysis period. We expect that the probability of engagement of a receiving user will increase with the number of friends that the receiver has, who are already engaged with the artist, as a consequence of either social influence, or homophily in the social network. 

We group all receivers who have the same number of friends, and track the probability of engagement as a function of the number of engaged friends of the receiver. The reason for holding constant the number of friends is to control for the absolute number of  friends, as well as the fraction who are engaged. The results are shown in Fig.~\ref{fig:pfan} (b) for 2,3 and 5 friends. The shaded area gives the $95\%$ confidence interval. In all cases, the probability of engagement increases with the number of engaged friends, supporting the hypothesis that social cohesion increases the probability of engagement. 

It is possible that this effect that we observe is related to the artist popularity. In particular, the fraction of listener friends who are engaged with an artist may simply be a proxy for popularity, as these artists have more engagement overall. To disentangle these effects, we separate out the artist according to their popularity on Spotify, and plot the probability of engagement as a function of number of engaged friends. The results are shown in Fig.~\ref{fig:pfan} (c). The probability of engagement increases with the number of engaged friends, even when accounting for artist popularity. More interestingly, we notice that this effect is more pronounced for the less popular artists, perhaps indicating that such social effects play a more important role in music discovery for lesser known artists. 

Theses measures are reminiscent of those used to investigate complex contagion in the spread of social behaviours. Complex contagion as defined in \cite{centola2007complex} is a model for how behaviours spread over social networks, with probability of infection increasing with the number of contacts. Evidence for this social effect of contagion has been found in empirical data from the spreading of campaign donations \cite{traag2016complex}, exercising \cite{aral2017exercise}, and marketing recommendations \cite{leskovec2007dynamics}. Our dataset does not contain evidence of multiple contacts; users are rarely sent a link to the same artist by multiple people. As in \cite{vanderweele2013social}, it can be difficult to disentangle the evidence of complex contagion and homophily, as both mechanisms produce similar results. However, our results on the relationship between social cohesion and engagement suggest that these social mechanisms could help explain our observations.


These mechanisms are typically related to clustering in the network. For this reason, we also experimented with the relationship between engagement and the receiver's neighbourhood clustering coefficient, and the edge overlap between the pair of users. The receiver's clustering coefficient is the proportion of friends who are also friends with each other, and higher clustering has showed to be important for complex contagion spread~\cite{centola2010spread}. The edge overlap as defined in \cite{mattie2021edge} is the proportion of friends shared by the sender and receiver. These measures were calculated by taking the unweighted version of the social network. Neither of these measures had a significant relationship with engagement probability. This may be because the Spotify social network is a partial representation of the real underlying social network which is unavailable to us.

\subsection{Predicting Post-share Engagement}

We would also like to evaluate the insights from H1-3 together with other contextual factors which might influence short-term engagement following a share. To this end, we design a supervised binary classification task, using data available at share-time to predict engagement in the week following the share. This allows us to elucidate further the social and contextual factors which explain whether a shared track will result in a successful person-to-person music recommendation.

\begin{table*}[ht]
\centering
\begin{tabularx}{\linewidth}{|c||X|}
\hline
    \textbf{Feature} & \textbf{Description} \\
    \hline
\hline
    \textbf{Social Tie Strength (ST)} & \\
    \hline
Sum of social interactions & Sum of the number of social interactions between the sender ($j$) and receiver ($i$) prior to share-time. Includes collaborative playlist additions, social listening sessions, and link shares between the two users in either direction. Sum of social interactions = $L_{s;ij} + L_{c;ij} + L_{l;ij} + L_{l;ji}$. 

\\ \hline

Direct link share & Boolean, True if the social media application type used to share the link was direct (e.g. Whatsapp), as opposed to broadcast or unknown. \\ \hline

Reciprocal link sharing & Boolean, True if receiver has ever shared a link with the sender prior to share. This is equivalent to $L_{l;ij} > 0$. \\ \hline

\textbf{Social Network Degrees (SN)} & \\
    \hline

Receiver link share in-degree & Number of link shares received by the receiver ($i)$ (from anyone) prior to share-time. This is equivalent to $\sum_{k} L_{l;ki}$ \\ \hline

Receiver link share out-degree & Number of link shares sent by receiver ($i$) prior to share-time (to anyone). This is equivalent to $\sum_{k} L_{l;ik}$.  \\ \hline

Sender link share out-degree & Number of link shares sent by the sender ($j$) before the share-time (to anyone). This is equivalent to $\sum_{k} L_{l;jk}$. \\ \hline

\textbf{Social Cohesion (SC)} &  \\
    \hline
    Fraction of engaged friends & ``Friend" is defined as someone with whom the receiver has a
reciprocal relationship in the social network prior to share-time. A friend is considered engaged if $E_{u, \alpha}(t_0, 180) > 180$, where $u$ is the friend user, $\alpha$ is the shared artist and $t_0$ is the start of the analysis period. \\ \hline

\textbf{Taste Similarity (TS)} & \\
    \hline
    sender-receiver cosine
similarity & Cosine similarity between receiver and sender. \\ \hline

receiver-track cosine
similarity & Cosine similarity between receiver and track. \\ \hline

\textbf{Track Context (TC)} & \\
    \hline

Artist popularity at
share-time & Rank of the primary artist in Spotify by number of streams at the time of share. \\ \hline

Release age at share-time & Time elapsed between album release time and share-time (seconds). \\ \hline

\textbf{Sender-Artist Engagement (SA)} &  \\
    \hline

Sender-artist engagement 7 days prior & Sender-artist engagement aggregated over 7 days prior to share-time.  \\ \hline

    \textbf{Receiver Platform Usage (PU)} & \\
    \hline
    is subscriber & Whether the user was a paid subscriber to Spotify when the share was opened. \\ \hline

receiver streaming 7 days prior & Number of hours of streamed music in the previous 7 days to opening the share. \\ \hline

receiver day on platform & Number of days passed since the user registered to the platform to the day the share was opened. \\ \hline

\end{tabularx}
\caption{Model features}
\label{features_described}
\end{table*}

\subsubsection{Features}
We describe the set of features used to inform the binary prediction problem in Table \ref{features_described}. They are separated
into groups according to the hypothesis we are testing.  Social Tie Strength (ST) captures the strength of the relationship between the sender and receiver, measured by the number of times they have previously interacted on the platform, the reciprocity of their relationship and the type of application used to share the link. This set of features is informed by our results in the Tie Strength Effects section. The Social Network Degrees (SN) measures describe the role of the sender and receiver in the social network in terms of their degrees. The Social Cohesion measures (SC) describe the popularity of the artist amongst the receiver's social network connections, and allows us to capture the social effects of engagement probability. This feature is informed by our results on Social Cohesion. The Taste Similarity (TS) features describe the music taste similarity between the sender and receiver, and between the receiver and the shared track. These capture different effects; the sender-receiver cosine similarity gives the similarity between the listening histories of the interacting users. The receiver-track cosine similarity captures the level of targeting; or how well does the sender know the receiver's taste. This feature set is informed by our results in the Music Taste Similarity section. The sender-artist engagement (SA) describes the sender's level of engagement with the artist prior to the share. We also include information about the Track Context (TC), or the track album age and popularity of the artist on Spotify. This allows us to control for the general popularity of the shared track, which may reflect its inherent quality, causing a higher probability of engagement. Lastly, we include features that describe the receiving user's activity levels on Spotify, calling this group the Receiver Platform Usage (PU) measures. This group of features are included because the engagement threshold is dependent on the receiver's Spotify usage. For users that are not active Spotify users, it may be impossible to meet the engagement threshold, even if they appreciated the shared track.  These features provide information about the receiver, including whether they use the paid subscriber version of Spotify (as opposed to the free version), how much music they streamed in the 7 days prior to the share, and the number of days that they spent on the platform prior to the share-time. 

\subsubsection{Model}
We train a Random Forest binary classification model using the scikit-learn python framework \cite{scikit-learn} on the full set of features and data ($\sim$2.72M link sharing events) using a randomised hyper-parameter search with 5-fold cross-validation and 500 fits varying the parameters of the model. We find that the best model performance is obtained with 300 estimators, and a maximum tree depth of 60 resulting in a ROC-AUC of 0.73 (see Table \ref{ablation_test} Full Model).

\subsubsection{Feature Importance}
\begin{figure}
\centering
\includegraphics[scale=0.32]{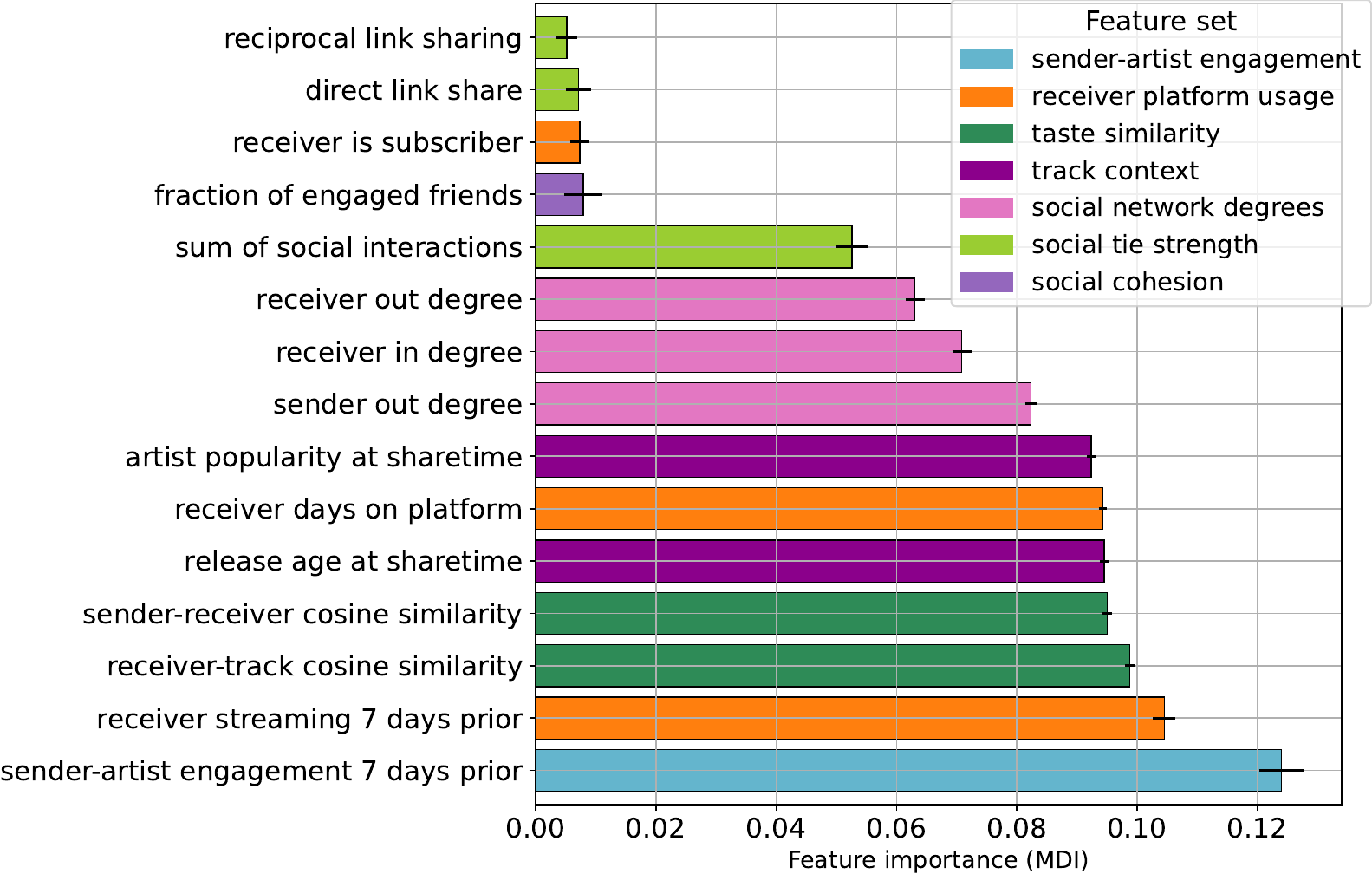}
\caption{Feature importances for the full model, measured using Mean Decrease in Impurity (MDI). Bars are coloured according to their feature set membership.}
\label{fig:mdi}
\end{figure}

From the model, we calculate the feature importance using Mean Decrease in Impurity (MDI). The feature importances are given in Fig.~\ref{fig:mdi}. The taste similarity features play an important role, as well as the receiver-platform engagement, sender-artist engagement, and various social features from the social network.  
Some of the features had surprisingly low importance in the full model. For example, the results in the Tie Strength and Social Cohesion sections indicated that the number of engaged friends, social media application type and reciprocal link sharing were important predictors of engagement, but these features do not appear as important for the full Random Forest classifier. This is likely because the information contained in these features can be ascertained using a combination of other features. For example, the model can determine that the number of engaged friends is greater than 0 using the sharer-artist engagement for the week prior to the share, and the number of previous shares between the receiver and the sender to determine if the relationship is reciprocal.

\begin{table*}[ht]
\centering
\begin{tabularx}{\linewidth}{|c|X|X|X|X|X|}\hline
    \textbf{Feature Set} & \textbf{ROC-AUC} & \textbf{Precision} & \textbf{Recall} & \textbf{Avg Precision}
    \\ \hline
    Social Cohesion (SC) &  0.5144 $\pm$ 0.0001 & \textbf{0.8779} $\pm$ 0.0004 & 0.0329 $\pm$ 0.0003 & 0.4868 $\pm$ 0.0003	
    \\ \hline
     Social Tie Strength (ST) &  0.5327 $\pm$ 0.0003 & 0.4997 $\pm$ 0.0002 & 0.1561 $\pm$ 0.0001 & 0.4933 $\pm$ 0.0003
    \\ \hline
    Social Network Degrees (SN) &  0.5469 $\pm$ 0.0002 & 0.5181 $\pm$ 0.0004 & 0.4091 $\pm$ 0.0005 & 0.5235 $\pm$ 0.0003
    \\ \hline
     Track Context (TC) &  0.5731 $\pm$ 0.0001 & 0.5356 $\pm$ 0.0003 
 & 0.5497 $\pm$ 0.0002 & 0.5365 $\pm$ 0.0003
    \\ \hline
    Taste Similarity (TS) &  0.5847 $\pm$ 0.0004 & 0.5346 $\pm$ 0.0005 & 0.5088 $\pm$ 0.0004 & 0.5935 $\pm$ 0.0005
    \\ \hline
    Receiver Platform Usage (PU) &  0.6102 $\pm$ 0.0005 & 0.5536 $\pm$ 0.0006 & 0.5573 $\pm$ 0.0004 & 0.6018 $\pm$ 0.0007 
    \\ \hline
    Sender-Artist Engagement (SA) &  0.6954 $\pm$ 0.0001 & 0.6487 $\pm$ 0.0005 & 0.5737 $\pm$ 0.0003 & 0.6634 $\pm$ 0.0003
    \\ \hline
    Full Model &  \textbf{0.7336} $\pm$ 0.0003 & 0.6734 $\pm$ 0.0004 & \textbf{0.5874} $\pm$ 0.0003 & \textbf{0.7498} $\pm$ 0.0004
    \\ \hline
\end{tabularx}
\caption{Feature Set Isolation Test Results. All metrics are reported with their standard error.}
\label{ablation_test}
\end{table*}

Feature importance measures are also impacted by feature correlations. Unsurprisingly, the taste similarity features are highly correlated with a 0.42 Pearson correlation ($p<0.01$). Several of the social network features are also correlated. For example, the sum of social interactions and the receiver in-degree are correlated with a Pearson correlation of 0.48 ($p<0.01$). The sum of social interactions is related to the amount that the receiver shares in general. Likewise, the sender out-degree is correlated with the sum of social interactions, with a correlation of 0.32 ($p<0.01$). This explains why the MDI scores for some features can be lower than expected, with the model taking one of the correlated features as redundant. To get a complete picture of the importance of various feature sets, we perform an isolation test described next.

\subsubsection{Feature Set Isolation Test}

We conduct a test to determine the relative importance of the various feature sets defined in Table \ref{features_described}. Instead of conducting an ablation test (removing one feature from the model at a time), which reveals little new information due to the variables being correlated, we conduct a feature set isolation test.  We train a Random Forest model (with the same hyper-parameters described above) but isolating the different groups of features, and compare the model performances to explore how predictive strength varies between them. We report the ROC-AUC, Precision, Recall and Average Precision. Since our dataset is not perfectly balanced, we use ROC-AUC score as an indicator of overall model performance. The results are summarised in Table \ref{ablation_test}, where the best performance is highlighted in bold. All standard errors of the metrics are within 0.0007, indicating stability across the 5-folds.

The sender-artist engagement (SA) and the receiver platform usage (PU) feature sets both perform well alone, with the highest ROC-AUC. As previously discussed, the sender-artist engagement could be a proxy for the strength of the sender's recommendation via text-based information that we do not have access to. The taste similarity features (TS), consisting of the sender-receiver cosine similarity and receiver-track cosine similarity, also perform reasonably well alone, as we might expect. However, taste similarity is not the highest performing feature set; this is likely due to the specific context of the share. The vectors capture the notion of taste, based on aggregated information over many contexts, which may have a smoothing effect on users with diverse taste profiles. The social cohesion feature (fraction of engaged friends) has the lowest ROC-AUC, but the highest precision. This feature is incredibly sparse, only available in approximately $1\%$ of the cases. For this reason it cannot have high recall which leads to lower overall performance. However, when the feature is available, it has the highest precision out of the feature sets, an effect we also observed in the section on Social Cohesion. Despite some sets of features performing better in isolation, all sets of features perform better than a random baseline (ROC-AUC = 0.5) and it is only when all features are combined that the best model performance is achieved.

   
   
    



    

    

    

    

    

    


\section{Conclusions and Future Work}
We have conducted an analysis of how link sharing between users on Spotify leads to new music discovery. 
First we determined that the Spotify social network exhibits homophily and that the probability of engagement is higher when the share occurs: 
 (1) between users with similar taste, when (2) the share is highly targeted to the the receiver's tastes, and (3) when the sender is more  engaged with the artist they shared (H1). Next, we demonstrated that the probability of engagement is higher along stronger and reciprocal ties, and is facilitated by intimate one-to-one modes of contact (H2). Through (H3),  we also explored how social cohesion affects the probability of engagement, demonstrating that users with a higher fraction of engaged friends will be more likely to become engaged with the shared artist. Finally, we built a classifier using a combination of social and contextual features, as well as individual attributes, to predict a new engagement following a person-to-person music recommendation, and elucidate which social and contextual factors contribute to the social discovery of music on Spotify. While the model we constructed performs well at predicting a new engagement, there may indeed be other important factors we are unable to capture with the data available to us, such  such as the type of relationship between the sender and receiver, the context in which the link was shared, and so on.

In this paper we have focused on short-term engagement resulting from a link share, one future research direction would be to investigate the lifetime of such engagement, and whether sharing can lead to more permanent behaviour changes, resulting in the receiver becoming a long-term listener, or even a fan, of the shared artist. Our work here was also limited to share events where the receiver opened and listened to the shared track. Another possible research direction would be to investigate which social and contextual factors affect the probability of a link being opened, a lower level of interest in the shared content. Since we have focused on the dynamics of pairwise interactions in the Spotify social network, other possible future directions include the study of higher-order group interactions, for instance to characterise taste homogeneity and the life-cycle of playlists.

This work contributes to a wider body of literature on how behaviours, tastes and opinions diffuse over a social network. Using explicit evidence of contact events together with rich contextual and social information, we have shed light on the dynamics of social music discovery in the age of online music streaming.

\section{Acknowledgements}
Special thanks to Glenn McDonald and Ward Ronan for their guidance and expertise on crucial datasets for this research.
We also thank  S. Elisha, R. Jones, D. Korkinof, A. McDowell, B. Regan, L. Vongsathorn for useful conversations and feedback on this work.  The work of R. Lambiotte was supported by EPSRC grants EP/V03474X/1 and EP/V013068/1. S.A. Babul was supported by EPSRC grant EP/W523781/1. This work was supported by The Alan Turing Institute’s Enrichment Scheme. The code for creating the multi-layer network in Fig.~\ref{fig:multilayer} comes from https://github.com/jkbren. 

\bibliography{ref.bib}

\appendix

\end{document}